\documentclass[twocolumn]{aastex61}

\usepackage{natbib}
\usepackage{graphicx}
\usepackage{hyperref}

\begin{document}

\newcommand{\LCO}{Las Cumbres Observatory, 6740 Cortona Drive, Suite 102, Goleta, CA 93117-5575, USA}
\newcommand{\UCSB}{Department of Physics, University of California, Santa Barbara, CA 93106-9530, USA}
\newcommand{\SAAO}{South African Astronomical Observatory, P.O. Box 9, Observatory 7935, Cape Town, South Africa}
\newcommand{\SALT}{Southern African Large Telescope Foundation, P.O. Box 9, Observatory 7935, Cape Town, South Africa}
\newcommand{\UCBPhysics}{Department of Physics, University of California, Berkeley, CA 94720, USA}
\newcommand{\UCBAstronomy}{Department of Astronomy and Theoretical Astrophysics Center, University of California, Berkeley, CA 94720, USA}
\newcommand{\LBLNuclear}{Nuclear Science Division, Lawrence Berkeley National Laboratory, 1 Cyclotron Road, Berkeley, CA 94720, USA}
\newcommand{\AMNH}{Department of Astrophysics, American Museum of Natural History, Central Park West and 79th Street, New York, NY 10024, USA}
\newcommand{\YNAO}{Yunnan Observatories, Chinese Academy of Sciences, 650011 Kunming, Yunnan Province, China}
\newcommand{\CAMS}{Center for Astronomical Mega-Science, Chinese Academy of Sciences, 20A Datun Road, Chaoyang District, 100012 Beijing, China}
\newcommand{\CAS}{Key Laboratory for the Structure and Evolution of Celestial Objects, Chinese Academy of Sciences, 650011 Kunming, China}
\newcommand{\Warsaw}{Warsaw University Astronomical Observatory, Al. Ujazdowskie 4, PL-00-478, Warszawa, Poland}
\newcommand{\Columbia}{Columbia Astrophysics Laboratory, Columbia University, New York, NY, 10027}
\newcommand{\Einstein}{Einstein Fellow}
\newcommand{\IOA}{Institute of Astronomy, University of Cambridge, Madingley Road, Cambridge CB3 0HA, UK}
\newcommand{\SUT}{Centre for Astrophysics and Supercomputing, Swinburne University of Technology, PO Box 218, H29, Hawthorn, VIC 3122, Australia}
\newcommand{\OzGrav}{The Australian Research Council Centre of Excellence for Gravitational Wave Discovery (OzGrav), Australia}
\newcommand{\CAASTRO}{The Australian Research Council Centre of Excellence for All-Sky Astrophysics (CAASTRO), Australia}
\newcommand{\AAO}{Australian Astronomical Observatory, 105 Delhi Road, North Ryde, NSW 2113, Australia}
\newcommand{\Swinburne}{Centre for Astrophysics and Supercomputing, Swinburne University of Technology, PO Box 218, H29, Hawthorn, VIC 3122, Australia}

\title{The Rapid Reddening and Featureless Optical Spectra of the optical counterpart of GW170817, AT 2017\MakeLowercase{gfo}, During the First Four Days}

\author{Curtis~McCully}
\altaffiliation{\href{mailto:cmccully@lco.global}{cmccully@lco.global}}
\affiliation{\LCO}
\affiliation{\UCSB}

\author{Daichi~Hiramatsu}
\affiliation{\LCO}
\affiliation{\UCSB}

\author{D.~Andrew~Howell}
\affiliation{\LCO}
\affiliation{\UCSB}

\author{Griffin~Hosseinzadeh}
\affiliation{\LCO}
\affiliation{\UCSB}

\author{Iair~Arcavi}
\affiliation{\LCO}
\affiliation{\UCSB}
\affiliation{\Einstein}

\author{Daniel~Kasen}
\affiliation{\UCBPhysics}
\affiliation{\UCBAstronomy}
\affiliation{\LBLNuclear}

\author{Jennifer~Barnes}
\affiliation{\Columbia}
\affiliation{\Einstein}

\author{Michael~M.~Shara}
\affiliation{\AMNH}
\affiliation{\IOA}

\author{Ted~B.~Williams}
\affiliation{\SAAO}

\author{Petri~V\"ais\"anen}
\affiliation{\SAAO}
\affiliation{\SALT}

\author{Stephen~B.~Potter}
\affiliation{\SAAO} 

\author{Encarni~Romero-Colmenero}
\affiliation{\SAAO}
\affiliation{\SALT}

\author{Steven~M.~Crawford}
\affiliation{\SAAO}
\affiliation{\SALT}

\author{David~A.~H.~Buckley}
\affiliation{\SAAO}
\affiliation{\SALT}

\author{Jeffery~Cooke}
\affiliation{\SUT}
\affiliation{\CAASTRO}
\affiliation{\OzGrav}

\author{Igor~Andreoni}
\affiliation{\SUT}
\affiliation{\OzGrav}
\affiliation{\AAO}

\author{Tyler~A.~Pritchard}
\affiliation{\Swinburne}

\author{Jirong~Mao}
\affiliation{\YNAO}
\affiliation{\CAMS}
\affiliation{\CAS}

\author{Mariusz~Gromadzki}
\affiliation{\Warsaw}

\author{Jamison~Burke}
\affiliation{\LCO}
\affiliation{\UCSB}

\begin{abstract}
We present the spectroscopic evolution of AT 2017gfo, the optical counterpart of the first binary neutron star (BNS) merger  detected by LIGO and Virgo,  GW170817. While models have long predicted that a BNS merger could produce a kilonova (KN), we have not been able to definitively test these models until now. From one day to four days after the merger, we took five spectra of AT 2017gfo before it faded away, which was possible because it was at a distance of only 39.5 Mpc in the galaxy NGC 4993. The spectra evolve from blue ($\sim6400$K) to red ($\sim3500$K) over the three days we observed. The spectra are relatively featureless --- some weak features exist in our latest spectrum, but they are likely due to the host galaxy.  However, a simple blackbody is not sufficient to explain our data: another source of luminosity or opacity is necessary. Predictions from simulations of KNe qualitatively match the observed spectroscopic evolution after two days past the merger, but underpredict the blue flux in our earliest spectrum. From our best-fit models, we infer that AT 2017gfo had an ejecta mass of $0.03M_\odot$, high ejecta velocities of $0.3c$, and a low mass fraction $\sim10^{-4}$ of high-opacity lanthanides and actinides. One possible explanation for the early excess of blue flux is that the outer ejecta is lanthanide-poor, while the inner ejecta has a higher abundance of high-opacity material. With the discovery and follow-up of this unique transient, combining gravitational-wave and electromagnetic astronomy, we have arrived in the  multi-messenger era.
\end{abstract}

\section{Introduction}\label{sec:introduction}

Simulations have predicted that as neutron stars (NSs) inspiral, some matter is tidally disrupted and is ejected \citep{Rosswog99, Goriely11}.  These could produce optical/IR emission, termed a ``kilonova'' (KN) or ``macronova'' (\citealt{Li98}, see \citealt{TanakaKNe, MetzgerKNe} for recent reviews). Models of  BNS mergers generally predict a few hundredths of a solar mass of ejecta with very high velocities of $0.1c-0.3c$ (30,000--90,000 km s$^{-1}$; \citealt{Bauswein13,Hotokezaka13, Kyutoku15, Sekiguchi16}). The resulting optical transients are expected to be fast, rising and fading over a few days, with a peak luminosity of $10^{40}-10^{41}$ erg s$^{-1}$ \citep{Metzger10, Barnes16, Tanaka17}. A key uncertainty of KN models is the composition of the ejecta, in particular, the content of lanthanides and actinides. These species have a high opacity, orders of magnitude larger than that of SNe ejecta (which is dominated by Fe-group elements; \citealt{Kasen13}), but are only produced by the $r$-process under very neutron-rich conditions \citep[e.g.][]{Lattimer74, Lattimer76, Lippuner15}. If the abundance of lanthanides/actinides in the ejecta is high, the KN emission is expected to peak in the red and into the infrared (IR; e.g.~\citealt{Barnes13, Barnes16}). However, if the ejecta is less neutron-rich, then the KN will be lanthanide-poor, and therefore blue \citep{Metzger14}. 

Short gamma ray bursts (GRBs) are also thought to be associated with BNS mergers based on time-scale and host galaxy arguments (\citealt{Berger14_GRB_Review} and references therein) and were the first targets for KN searches \citep[e.g.][]{Perley09, Yang15, Jin16}. GRB 130603B ($z=0.3568$; \citealt{Postigo14}) produced two types of optical emission: shortly after the GRB, an afterglow was detected by the \textit{Swift} satellite \citep{Evans13}. Afterglows are likely produced by a shock interacting with the surrounding medium producing emission from the X-ray to the radio. The optical afterglow of GRB 130603B rose and faded in less than one day \citep{Cucchiara13}, but excess in the IR remained, which was interpreted as light from a KN \citep{Berger13, Tanvir13}.

\begin{deluxetable*}{ccccccc}
\tablecaption{Spectroscopic observation log of the optical counterpart of GW170817, AT 2017gfo\label{tab:spectra}}
\tablehead{\colhead{UT Date/Time} & \colhead{Phase} & \colhead{Exposure Time} & \colhead{Telescope} & \colhead{Instrument} & \colhead{Resolution} & \colhead{Wavelength Coverage}} 
\startdata
2017 Aug 18 17:07:20 & $+1.18$ days & 433 s & SALT & RSS & 300 & 3600-8000 \AA \\
2017 Aug 19 08:36:22 & $+1.81$ days & 3600 s & LCO FTS & FLOYDS & 700 & 5500-9250 \AA \\
2017 Aug 19 16:58:32 & $+2.16$ days & 716 s & SALT & RSS & 300 & 3600-8000 \AA \\
2017 Aug 20 01:01:54 & $+2.49$ days & B600:128 s/R400:763 s & Gemini-South & GMOS & 600/400 & 5500-9500 \AA \\
2017 Aug 21 00:16:09 & $+3.45$ days & B600:1440 s/R400:1440 s & Gemini-South & GMOS & 600/400 & 4500-9500 \AA \\
\enddata
\tablecomments{Phase is given in rest-frame days.}
\end{deluxetable*}

\section{GW170817/GRB 170817A}
Short gamma ray burst GRB 170817A, was detected by the \textit{Fermi} satellite on 2017 August 17 12:41:06 UTC \citep{FermiLVC, Goldstein17}, nearly coincident in time with a signal ending about 2 s earlier by the Hanford detector of the Laser Interferometer Gravitational-wave Observatory (LIGO; \citealt{LIGO}), ultimately named GW170817. LIGO Livingston and Virgo \citep{Virgo} data were later combined to produce an estimate of the 3-dimensional position of the gravitational-wave (GW) source: a $\sim 32$ deg$^2$ region on the sky at a distance of $40 \pm 8$ Mpc \citep{GWPRL}. The small region of sky afforded by using all three GW detectors and the nearby search volume made this a prime candidate for optical follow-up to search for an associated KN.

We triggered a search for an optical counterpart using the Las Cumbres Observatory \citep[LCO;][]{Brown13} global network of 20 robotic telescopes (see \citealt{ArcaviStrategy} for a full description of our strategy).  We imaged a list of galaxies in the search volume, and in the fifth galaxy on the list, NGC 4993, detected a candidate optical counterpart at $\alpha_{2000}=$13$^{\textrm{h}}$09$^{\textrm{m}}$48\fs07 and decl., $\delta_{2000}= -23^{\circ}22\arcmin53\farcs7$ \citep{ArcaviGW}.  Within a span of 42 minutes, six collaborations had independently imaged the same target, before the discovery announcement: the Swope Supernova Survey (SSS) and One-Meter Two-Hemisphere (1M2H) Collaboration \citep{CoulterLVC, SSSGW}; Distance Less Than 40 Mpc (DLT40; \citealt{DLT40GW, ValentiLVC}); the Dark Energy Survey (DES) GW Community \citep{BergerLVC, DECamGW}; the LCO GW Followup Team \citep{ArcaviLVC, ArcaviGW}; the Mobile Astronomical System of TElescope Robots (MASTER) Collaboration \citep{MasterLVC, MASTERGW}; and the VIsta Near-infraRed Observations Unveiling Gravitational wave Events (VINROUGE) Collaboration \citep{VISTALVC, VISTAGW}.  The transient was first imaged and first reported by \citet{CoulterLVC} as SSS17a on behalf of the SSS/1M2H collaboration. It was next imaged and later reported by the DLT40 as DLT17ck and eventually was given the IAU name AT 2017gfo. The transient's host galaxy is NGC 4993, an S0 galaxy at $z=0.009727$ at 39.5 Mpc \citep{NED_Distance}; see \citet{Discovery} for an overview of the discovery. 

We obtained five spectroscopic follow-up observations of AT 2017gfo taken between +1.18 and +3.5 rest-frame days after the merger GW170817, roughly every 12 hr. We obtained our first and third spectra of AT 2017gfo from the Southern African Large Telescope (SALT; \citealt{Buckley06}) using the Robert Stobie Spectrograph (RSS; \citealt{RSS}) under Directors Discretionary Time (program 2017-1-DDT-009). Our second spectrum was obtained using the FLOYDS spectrograph on the Faulkes Telescope South (FTS) in the LCO network. The final two spectra reported here were taken on the Gemini-South telescope and were taken under a joint agreement between GS-2017B-Q-14 (PI: Howell) and GS-2017B-DD-1 (PI: Singer), which included the sharing of Gemini spectra and IR photometry obtained by either group. The second spectrum was triggered under GS-2017B-Q-30 (PI: Troja) and was shared with the Singer/Howell consortium. See also \citet{Kasliwal} and \citet{Troja}.

\section{Observations}\label{sec:observations}
A summary of our spectroscopic observations is given in Table \ref{tab:spectra}.

Our first spectrum of AT 2017gfo was taken using RSS on SALT at 17:07:19.703 UTC on 2017 August 18. The spectrum was taken using the PG300 grating (with a resolution of $\sim 300$) in the 5\fdg75 grating angle with the 2\arcsec slit. This gave us a wavelength coverage of 3600-8000 {\AA}. The observation reported here has an exposure time of 433 s and was obtained at an airmass of 1.39. Observations were taken in early twilight due to the location of the source and the visibility limitations of SALT, and these observations are contaminated with a high sky background. 

Data reduction was carried out with the PySALT package \citep{Crawford10} including removal of basic CCD characteristics, cosmic-ray cleaning, wavelength calibration, and relative flux calibration. The extraction of the flux was performed by simultaneously fitting the flux from the host galaxy, the atmospheric sky lines, and the source spectra using the \textit{astropy.modeling} package \citep{Astropy}. We used observations of the spectrophotometric standard, EG21, from the end of the night of 2017 August 18 for flux calibration.

SALT's primary mirror is fixed during an observation, and a moving instrument package at prime focus tracks the target in the focal plane; this results in a limited window of visibility for any particular object and time-varying sensitivity during the observation. The optical design of SALT has an 11 m diameter entrance pupil, but not all of the pupil is covered by the primary mirror array, i.e., the pupil is underfilled. As the prime focus tracks an object, the center of the pupil migrates across the primary mirror array, leading to a varying effective collecting area as a function of track time; the effective collecting area varies between 7 and 9 m \citep{Stobie00, Buckley06}. Because of the changing pupil during SALT observations, only relative flux calibration can be achieved.

 The small variations we observe in the first SALT spectrum in the blue, $<5000$ \AA\ are likely regions of increased noise due to the high sky background. The high sky background and low S/N also led to some systematic uncertainty in the flux calibration in the blue.

Our next spectrum was taken on LCO FTS with the FLOYDS spectrograph. The target was only visible for a short time, setting only a little more than an hour past twilight. On 2017 August 19, we obtained an hour-long spectroscopic exposure. The source was fading rapidly at this point, so the signal-to-noise ratio (S/N) of the spectrum was very low. However, a trace was visible, and we extracted the spectrum using the FLOYDS pipeline\footnote{\url{https://github.com/svalenti/FLOYDS_pipeline}} \citep{Valenti_FLOYDS}. Because the S/N was low, we then warped the spectrum to match the \textit{gri} photometry (see \citealt{ArcaviGW} for a full description of the photometric follow-up of this event) to ensure that our characterization of the spectral energy distribution (SED) is accurate.

\begin{figure*}
    \centering
    \includegraphics[width=\textwidth]{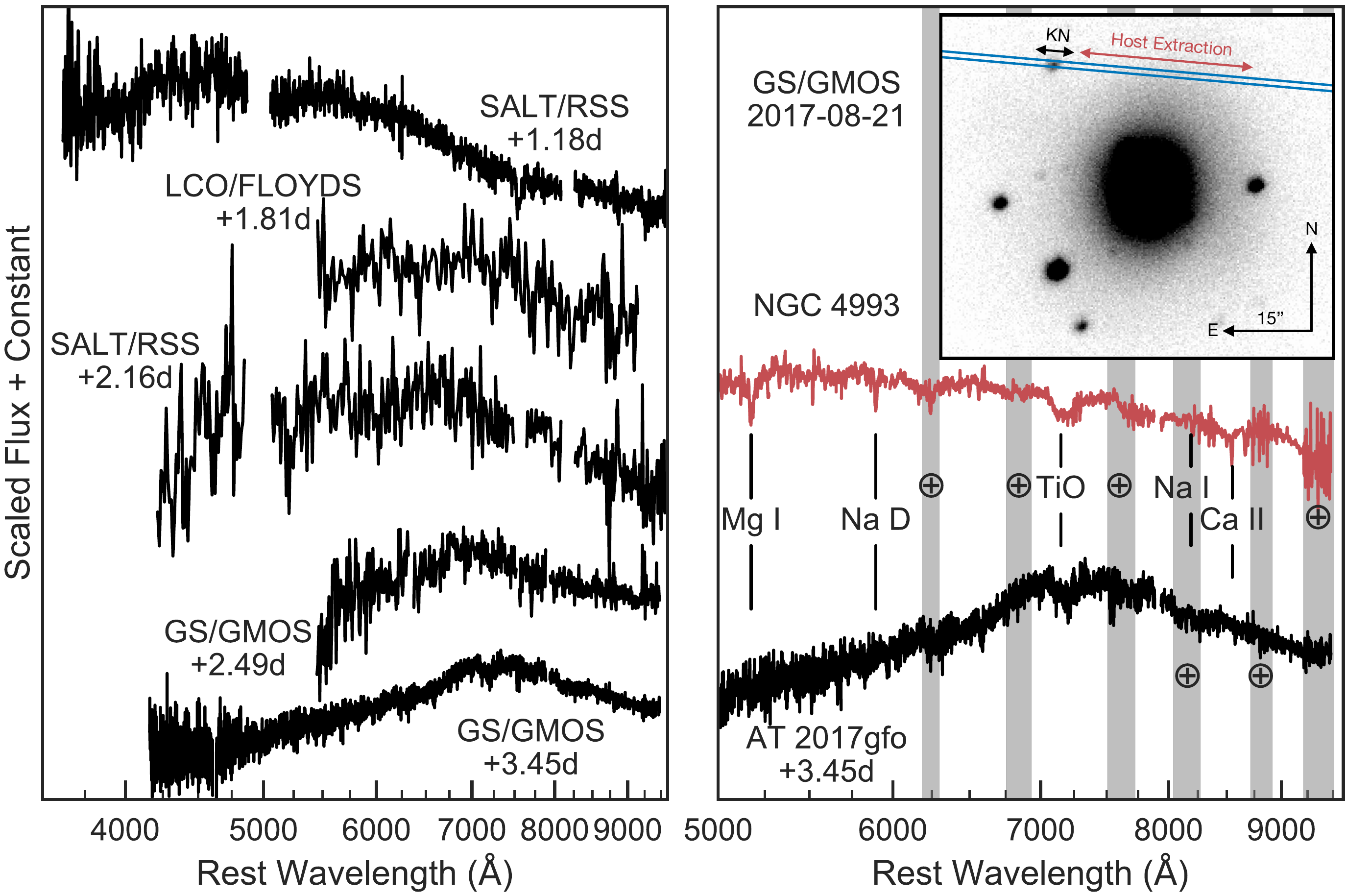}
    \caption{Spectroscopic evolution of the KN optical counterpart. The FLOYDS, the second SALT, and the first Gemini spectra are low S/N, so we show binned data. The S/N in the blue of the first SALT spectrum is low, so any apparent features below $\sim 5000$ \AA\ are likely noise. The spectra evolve from blue to red over less than 2.5 days: the peak of the emission in the SALT spectrum is at a rest wavelength of $\sim 4500$ \AA, while the peak is at $\sim 7500$ \AA\ in our latest spectrum. The spectra are dominated by continuum and are nearly featureless. The right panel shows a comparison of the spectrum of the KN candidate and its host. Both spectra were extracted from the the GS spectrum taken on 2017 August 21 as illustrated in the inset. The top spectrum is an extraction of the host galaxy, NGC 4993, and the bottom is an extraction of the KN candidate. Many of the features visible in the KN candidate spectrum also appear in the host spectrum. We have labeled some of the strongest features from the host and have marked regions that have large telluric correction in gray with the $\earth$ symbol. We interpret this to mean that the KN candidate is intrinsically featureless, while the features we observe in the spectrum are due to absorption by interstellar material in the host galaxy.}
    \label{fig:spectra}
\end{figure*}

A second SALT/RSS spectrum was taken at 16:58:32 UTC on 2017 August 19 with the same setting as the first, with an exposure time of 716 s. The trace of the transient was still clearly visible in the twilight though the resulting spectrum is of low signal-to-noise ratio (S/N).  A third spectrum was taken on the following night but it was too contaminated by background to be useful.

 The first spectrum from Gemini used the B600 and R400 gratings with an exposure time of 128 s at a central wavelength of 520 nm and 763 s at a central wavelength of 720 nm, respectively. These observations were taken at high airmass ($\sim 2.9$) on 2017 August 20 UTC. This spectrum also had low S/N, but like in the FLOYDS spectrum, a trace was visible.
 
 The final spectrum we obtained was using GMOS on 2017 August 21 (airmass $\sim 1.5$). For this spectrum, we obtained $4\times300$ s exposures in both the B600 and R400 (with the same central wavelengths from our first GMOS spectrum). This spectrum was much higher S/N than our previous two. The Gemini data were reduced using the standard techniques using a combination of the Gemini-IRAF\footnote{IRAF is distributed by the National Optical Astronomy Observatory, which is operated by the Association of Universities for Research in Astronomy (AURA) under a cooperative agreement with the National Science Foundation.} and custom procedures written in Python (see \url{https://github.com/cmccully/lcogtgemini}). Because these data were taken at high airmass and this is a new class of optical transient, special care was taken with the flux calibration (sensitivity function) and telluric correction. We obtained observations of the standard star, EG 274, after both of our Gemini-South observations. We adopt the model from \citet{Moehler14} of EG 274 for our flux calibration. In the red, using the R400 grating, the spectrum cut off at $\sim 9500$ \AA\ in the middle of a telluric absorption feature. To account for this, we combined the sensitivity function with one derived from observations taken of EG 274 using the R400 grating, but with a central wavelength of 8000 {\AA}. The telluric correction is then derived from the standard star observation taken soon after the science spectrum. 

We corrected for Milky Way reddening using the \citet{Cardelli} extinction law. We adopted $R_V = 3.1$ and $A_V = 0.329$ from \citet{Schlafly}, which was obtained using the Astroquery package \citep{astroquery}.

These spectra will be available on the Weizmann Interactive Supernova data REPository (WISeREP; \citealt{Yaron12})

\begin{figure}
    \centering
    \includegraphics[width=\columnwidth]{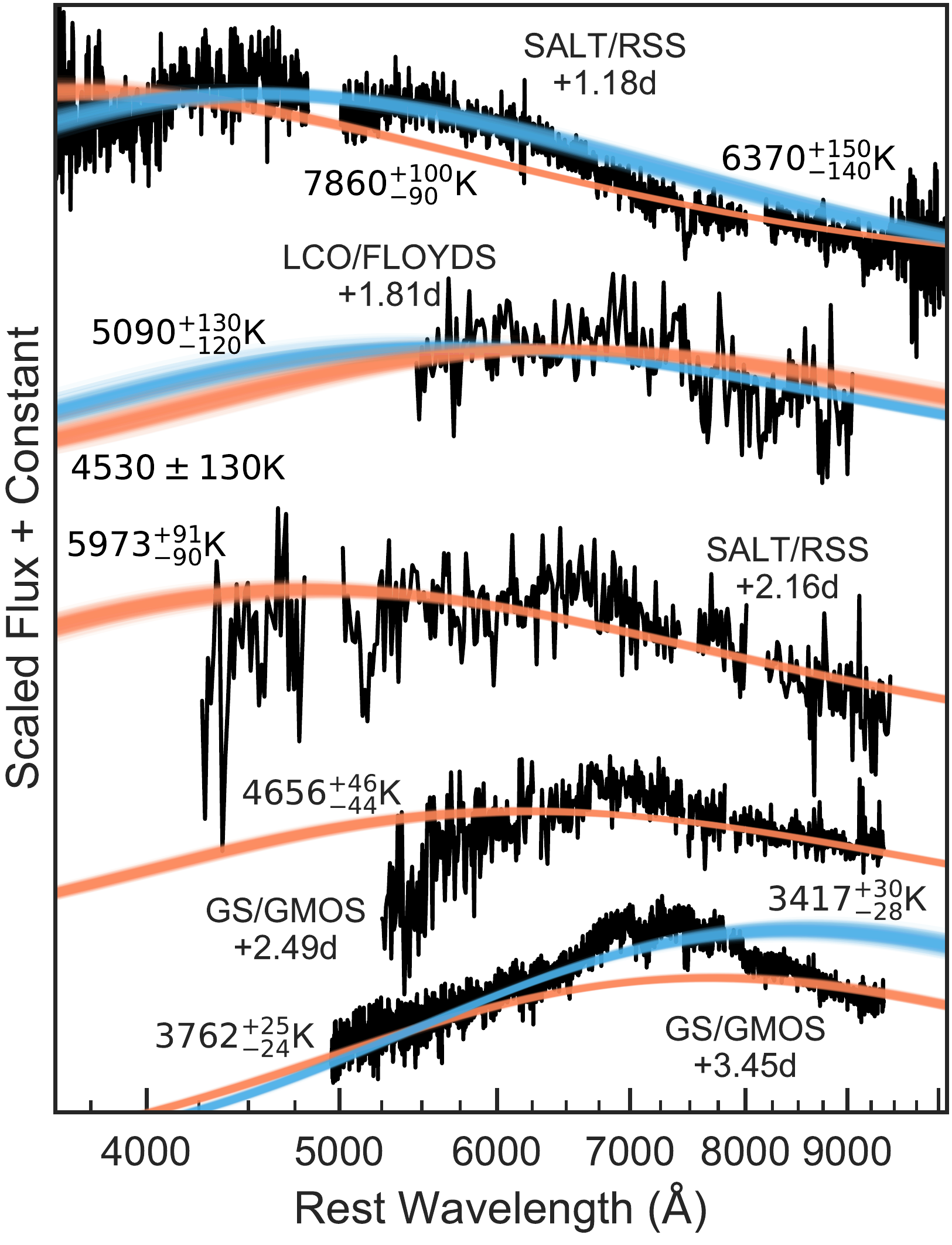}
    \caption{Blackbody fits to the spectra of the KN candidate. MCMC samples of our blackbody model are shown by colored lines. The best-fit models evolve from $\sim6000$K to $3762^{+25}_{-24}$K over less than 2.5 days. For all of the blackbody fits, the parameter space is multi-modal. This is especially true for the first SALT spectrum. The ``best fit'' from MCMC is $7360^{+100}_{-90}$K. However, the concavity of this model is wrong. We show a $6370^{+150}_{-140}$K blackbody in blue: this is a much better fit to the peak, even though the $\chi^{2}$ is worse than the higher-temperature fits. We also have shown blackbody models for two temperatures, $5090^{+130}_{-120}$ K and $4530 \pm 130$K (blue and orange, respectively) compared to the FLOYDS spectrum. The higher-temperature fit is preferred from the full dataset, but if we exclude telluric regions that may be over corrected, the lower-temperature solution is favored. In both of the latest two spectra, the best-fit model underpredicts the peak flux. In the final spectrum, if only the region blueward of the peak at $\sim 7800$ \AA\ is included, then we obtain a best-fit temperature of $3417^{+30}_{-29}$ K. This then overpredicts the red flux that we observe while still underpredicting the peak. While these spectra are featureless, a simple radiating blackbody is not a good fit to any of the spectra.}
    \label{fig:bb_fit}
\end{figure}

\section{Analysis}\label{sec:analysis}
The spectroscopic time series of the KN candidate is shown in the left panel of Figure \ref{fig:spectra}. A comparison with the host galaxy spectrum extracted from the same observation is shown in the right panel. 

The spectra of AT 2017gfo are mostly featureless. The only features visible are seen in the latest (and highest S/N) spectrum. However, those features are likely not intrinsic to the KN, but are instead due to absorption from the host galaxy. In the right panel of Figure \ref{fig:spectra}, we show the spectrum of the transient compared to the host. The strongest features in the spectrum of the transient match those we see in the host galaxy and are marked with the species that produce these absorption features in the spectra of elliptical galaxies \citep[e.g.][]{SparkeGalaxyBook}. During the spectral reduction, much of the host light was subtracted, so these features may be residuals from undersubtraction or could be due to absorption by interstellar material in the host galaxy.

\begin{figure}
    \centering
    \includegraphics[width=\columnwidth]{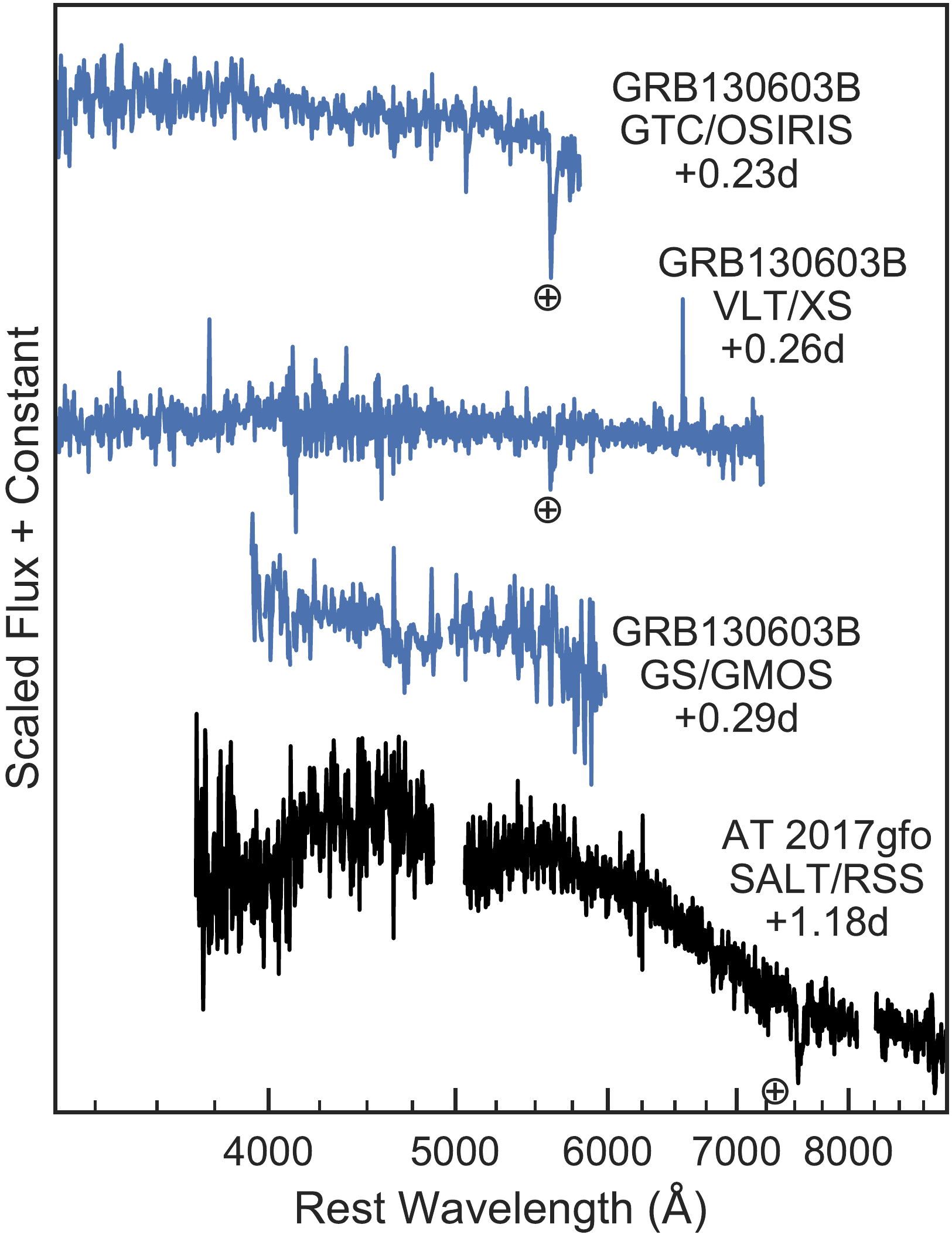}
    \caption{Spectra of the AT 2017gfo (black) compared to the spectra of afterglow from GRB 130603B (blue). Both the spectra of the afterglow of GRB 130603B and the spectrum of AT 2017gfo are featureless and have similar shapes. The GTC (top) and Gemini spectrum (third from the top) appear to have a turnover at the red end of the spectrum. This would be slightly bluer than the turnover in the SALT spectrum but only by a few hundred angstroms. The VLT (middle) spectrum of the afterglow of GRB 130603B has low S/N but is consistent with the other spectra.}
    \label{fig:grb_compare}
\end{figure}

\begin{figure*}
    \centering
    \includegraphics[width=\textwidth]{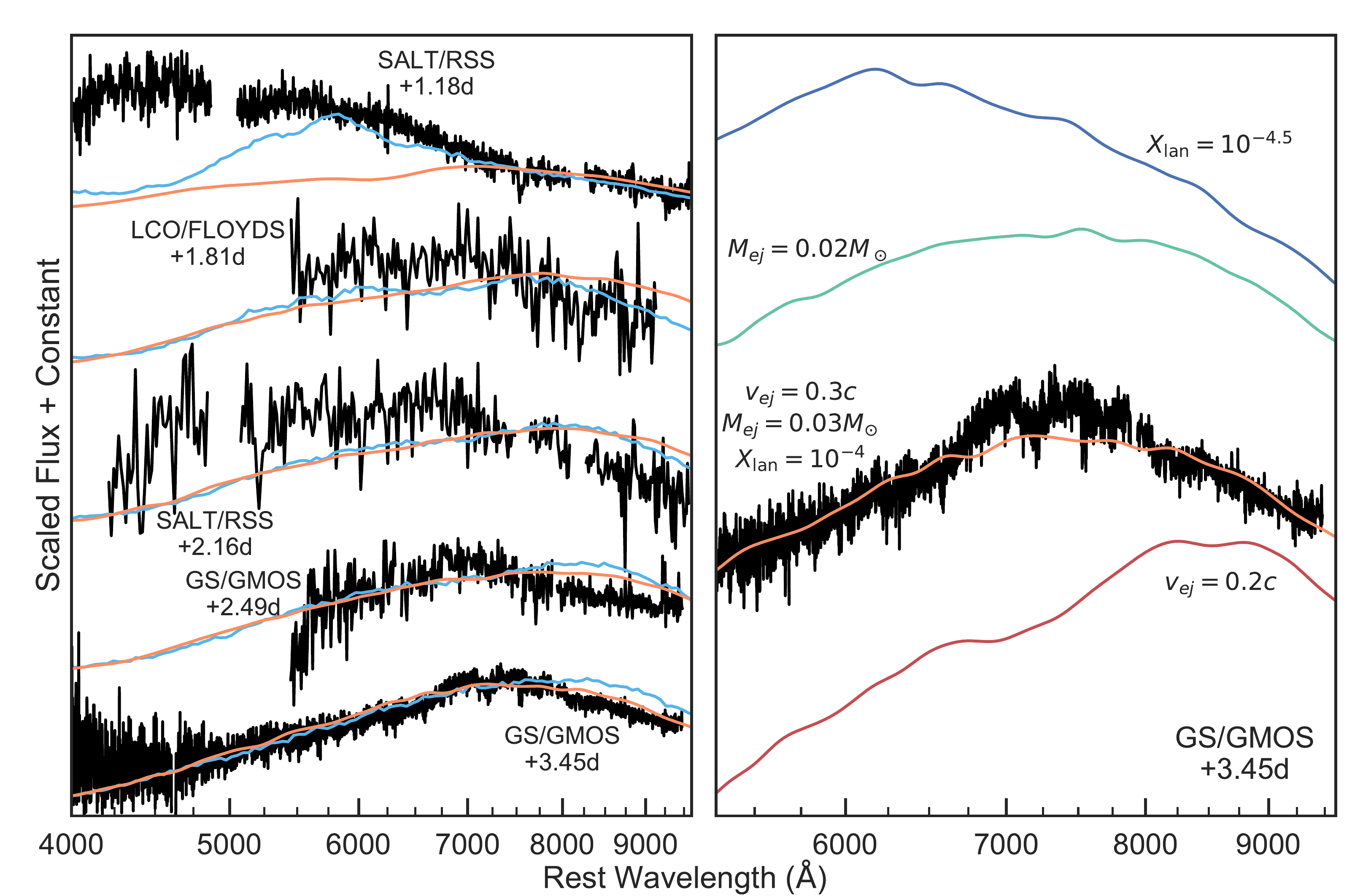}
    \caption{Comparisons of observed spectra to KNe models from \citet{Kasen17}. The left panel shows the spectroscopic time series as in Figure \ref{fig:spectra} and a model with a compositional gradient of lanthanides (blue). The outer parts of the ejecta have a lanthanide mass fraction of $10^{-6}$ while the inner ejecta has $10^{-4}$. This model has an ejecta mass $M_{ej} = 0.025 M_\odot$ and velocity $v_{ej} = 0.25c$. A constant-composition model with lanthanide fraction of $10^{-4}$, an ejecta velocity of $v_{ej}=0.3c$, and ejecta mass of $M_{ej} = 0.03 M_\odot$ is shown in orange. Overall, the KN models are much better fits than a single blackbody and qualitatively fit observed spectra after two days past the merger. However, at early times, roughly one day after merger, the models underpredict the blue flux we observe. The constant-composition model predicts even less blue flux than the compositional-gradient model. The right panel shows the best-fit model with a uniform composition (same as the orange line in the left panel). The other colored lines illustrate how differences in the  lanthanide fraction (blue, top), ejecta mass (green, second from the top), and ejecta velocity (red, bottom) manifest themselves in the models (the models are smoothed to remove numerical noise and offset for display purposes only). Each model has a single differing parameter (ejecta mass, velocity, or lanthanide mass fraction) than the best fit in orange, but the other parameters are the same. The top blue line shows the model spectrum for a lower lanthanide mass fraction of $10^{-4.5}$ which yields a bluer spectrum than is observed. The middle teal line shows the model spectrum for a lower ejecta mass of 0.02 $M_\odot$. In this wavelength range, the differences between these models are small. The higher ejecta mass causes the peak near $7500$ \AA\ to be slightly sharper. The bottom red line shows the model spectrum for a lower ejecta velocity of $0.2 c$. The decrease in ejecta velocity makes the spectrum redder at this phase of evolution.}
    \label{fig:model}
\end{figure*}

As the spectra are mostly featureless, the main observational constraint is the temperature. We fit a blackbody to each of the spectra using a Monte Carlo Markov Chain (MCMC) using the \texttt{emcee} package \citep{emcee}. The results of our temperature fits are in Figure \ref{fig:bb_fit}. We find the following best-fit temperatures: $t=1.18d$: $7860^{+100}_{-90}$ K; $t=1.81d$: $5090^{+130}_{-120}$ K; $t=2.16d$: $5973^{+91}_{-90}$ K; $t=2.49d$: $4656^{+46}_{-44}$ K; and $t=3.45d$: $3762^{+25}_{-24}$ K. The evolution from blue to red is consistent with theoretical predictions \citep[e.g.][]{Barnes16, Kasen17}, but in all of our blackbody fits, we find that the parameter space has several local minima. 

We find that the best-fit temperature for the first SALT spectrum is $7360^{+100}_{-90}$K, but even though this fit has the lowest $\chi^2$, it is not a good estimate of the temperature. If we only fit the the blue side of the spectrum $<6000$ \AA, then the best-fit model has a temperature of $6370^{+150}_{-140}$K. This matches the peak better, and is therefore likely a better estimate of the temperature.

In Figure \ref{fig:bb_fit}, we show two families of solutions that are consistent with the FLOYDS data (second spectrum from the top), one with temperature of $5090^{+130}_{-120}$ K and one with $4530 \pm 130$ K. When excluding telluric regions, the lower-temperature fit is preferred. Neither of the two latest spectra are well fit by a blackbody: the peak is under-predicted. If we only include the flux blueward of the peak at $\sim 7800$ \AA\ in our latest spectrum (bottom), we find a lower blackbody temperature of $3417^{+30}_{-28}$ K, but this overpredicts the observed flux in the red and still underpredicts the peak flux. We conclude that the spectra are not consistent with a single radiative blackbody.

Using two independent blackbodies could account for two possible emission components as suggested by some KN models \citep{Metzger10, Barnes13, Kasen13, Kasen15, Kasen17, Metzger14, Barnes16, Fernandez17, Rosswog17, Wollaeger17}. The addition of a second blackbody does not yield improved fits and therefore is not shown here. This implies that there is an extra source of opacity (or luminosity), consistent with the prediction of KN models. Our optical spectra primarily constrain the bluer emission from ejecta of low lanthanide abundance, while observations of spectra taken at infrared wavelengths would allow for strong constraints on the properties of a second, high-lanthanide ejecta component.

As mentioned above, the BNS merger, GW170817, had a corresponding short GRB, GRB 170817A. About half of short GRBs have observed corresponding afterglows \citep{Berger14_GRB_Review}, so in our earliest/bluest spectrum, there could be some contamination from an afterglow on top of the emission from the KN. At early times, AT 2017gfo had similar colors to previously discovered optical afterglows of short GRBs \citep{NichollLVC}. The afterglow of GRB 130603B was $\sim 4$ mag brighter than AT 2017gfo \citep{Tanvir13, ArcaviGW}. If there was an afterglow of GW170817/GRB 170817A, it was considerably weaker than the one observed from GRB 130603B. Also, the gamma rays detected from GRB 170817A were about three orders of magnitude dimmer than those from GRB 130603B \citep{Berger14_GRB_Review, Goldstein17, Discovery}, and no early X-ray flux was detected \citep{Evans21550}.

In Figure \ref{fig:grb_compare}, we compare spectra of AT 2017gfo with those of the afterglow of GRB 130603B, a previous KN candidate, from Gemini \citep{Cucchiara13}, X-Shooter (XS) on the Very Large Telescope (VLT), and OSIRIS on the Gran Telescopio CANARIAS (GTC; \citealt{Postigo14}). 

The OSIRIS/GTC spectrum had an exposure time of 900 s and was taken 7.4 hr after the GRB. The Gemini data were reprocessed using the same procedure as described in Section \ref{sec:observations}. The Gemini Archive included two observations of the afterglow of GRB 130603B, with the first of $2\times900$ s at 9.6 hr after the detection of the GRB (third from the top) and the second of 900 s, 31.2 hr after the GRB but the S/N was too low to be useful. The X-shooter spectrum was taken 8.6 hr after the detection of the GRB (middle).

The GTC and the first Gemini spectra of the afterglow of GRB 130603B (top and third down) have a slight change in the slope at $\sim 5750$ \AA\ (the VLT spectrum of the afterglow of GRB 130603B does not show the downturn seen in the SALT spectrum of AT 2017gfo, but is low S/N). This is similar to what is seen in the spectrum of AT 2017gfo, but the turnover is slightly redder for AT 2017gfo. The similarity might suggest that the first spectrum of AT 2017gfo from SALT has some contribution from an afterglow, even though the afterglow must have been much weaker than for GRB 130603B.

Models of KNe generally predict that the opacity will be driven by $r$-process elements; the optical emission is very sensitive to the abundance of lanthanides/actinides \citep[e.g.][]{Kasen13}. In Figure \ref{fig:model}, we compare our spectroscopic time series with the model predictions of \citet{Kasen17}. These models are spherically symmetric and assume homologous expansion. The density profile of the models is described by a broken power law: the density in the inner layers falls like $v^{-1}$ and more steeply, $v^{-10}$, in the outer layers. The transition between the power laws is set by the mass and kinetic energy \citep[see][]{Barnes13}. The radiation transport is calculated using the Sedona code \citep{Kasen06}. We consider two types of models: those with constant composition (constant lanthanide mass fraction) and models with a compositional gradient (the lanthanide mass fraction varies from the inner to the outer layers of the ejecta).

 At epochs $\gtrsim2$ days, the KN models show a significantly better match to the observed spectra (shown in Figure \ref{fig:model}) than do the blackbody fits (shown in Figure \ref{fig:bb_fit}). In particular, the KN models predict a more sharply “peaked” spectrum than a blackbody, with the flux falling off more sharply above and below the wavelength of maximum flux. This spectral shape is a consequence of the strong wavelength dependence of the line opacity that dominates the absorption and emission from KNe. That this peaked spectral shape better matches the observations than the shallower blackbody shape provides additional evidence for line-dominated emission as expected from a KN. On the whole, the KN spectra qualitatively match the observed spectra well, but at some wavelengths show deviations at the level of $\sim60\%$ for the second SALT spectrum, $\sim20\%$ for first Gemini spectrum, and $\sim10\%$ for the latest spectrum. The quantitative agreement could presumably be improved by modifying the model abundance gradient, which affects the time evolution of the spectral peak in the models. Such fine tuning of the model compositional structure has not, however, been attempted here.

In the left panel of Figure \ref{fig:model}, we compare our observed spectra to models from \citet{Kasen17}. In the left panel, we compare our observed spectra to two types of models: one with constant composition and one with a gradient in the abundance of lanthanides. The best-fit constant-composition model has an ejecta mass of 0.03 $M_\odot$ of ejecta, a high ejecta velocity of $0.3 c$, and a lanthanide mass fraction of $10^{-4}$. These parameters are similar to those found independently from the light curve analysis in \citet{ArcaviGW}. The high velocities of this model produce significant line blending that explains why the spectra appear to be featureless. 

The compositional-gradient model has a lanthanide mass fraction of $10^{-6}$ in the outer layers, but $10^{-4}$ in the inner layers, and has an ejecta mass $M_{ej} = 0.025 M_\odot$ with velocity $v_{ej} = 0.25c$. In the earliest spectrum of AT 2017gfo, both types of models significantly underpredict the blue flux we observe, consistent with what is found in \citet{ArcaviGW}. The compositional-gradient models are better fits than the models with a constant composition, but still underpredict the flux in the blue.

In the right panel of Figure \ref{fig:model}, we show the best-fit constant-composition model (the same parameters as the left panel; see above) and illustrate how changing specific parameters affects the models at a phase of +3.5 days. The top line shows the model when we decrease the lanthanide fraction to $10^{-4.5}$ which makes the model bluer. The next line shows the model spectrum for a lower ejecta mass of $0.02 M_\odot$. The difference in the ejecta mass does not change the observed spectrum much; the only difference is that peak is slightly sharper for a higher ejecta mass. The bottom line shows the model with lower ejecta velocities, $0.2c$: this has the effect of producing a redder spectrum than we observe. 

\section{Discussion and Conclusions}\label{sec:conclusions}
We have presented the spectroscopic evolution of AT 2017gfo, the optical counterpart of the first binary NS merger GW170817. The spectra evolve from the blue to red over about three days, though a simple radiative blackbody model is not sufficient to explain the spectra. There is likely some other source of opacity or luminosity causing the SED to differ from that of pure blackbody distribution.  

Generic KN models predict red, featureless spectra, consistent with the observations. We found that the best-fit models have high velocities and low lanthanide fractions. Models with lanthanides buried below the surface layers improved the fit in the blue, but a more complete parameter study is necessary to test if these models can sufficiently account for our observations. 

GW170817 was an extremely fortunate discovery: we did not expect for the first BNS merger to be discovered to be so close or to have the proper alignment to observe the associated short GRB \citep[e.g.][]{MetzgerKNe}. Beyond the opportunity to study an exciting new class of transients, the successful coordination of the LIGO/Virgo GW detectors and the electromagnetic observers makes this an exciting precedent for the future of extragalactic multi-messenger astronomy.

\acknowledgements
We thank the anonymous referee for their insightful comments and rapid response. We thank Antonio de Ugarte Postigo for sending us the GTC spectrum of GRB 130603B. We thank Tom Matheson for useful discussion and advice on reducing the Gemini spectrum. We thank Stefano Valenti and Dovi Poznanski for their work on the original proposal from LCO. We thank the LCO staff, specifically Mark Bowman and Mark Willis, and the Gemini staff, specifically Karleyne Silva and Laura Ferrarese, for their assistance with these observations. This work made use of the LCO network. C.M., G.H., and D.A.H. are supported by NSF grant AST-1313484." Support for I.A. and J.B. was provided by the National Aeronautics and Space Administration through Einstein Postdoctoral Fellowship Award Numbers PF6-170148 and PF7-180162, respectively issued by the \textit{Chandra X-ray Observatory} Center, which is operated by the Smithsonian Astrophysical Observatory for and on behalf of the National Aeronautics Space Administration under contract NAS8-03060. D.K. is supported in part by a Department of Energy (DOE) Early Career award DE-SC0008067, a DOE Office of Nuclear Physics award DE-SC0017616, and a DOE SciDAC award DE-SC0018297, and by the Director, Office of Energy Research, Office of High Energy and Nuclear Physics, Divisions of Nuclear Physics, of the U.S. Department of Energy under Contract No. DE-AC02-05CH11231. M.M.S. gratefully acknowledges the support of the late Paul Newman and the Newman’s Own Foundation. D.B., S.M.C., E.R.C., S.B.P., P.V., and T.W. acknowledge support from the South African National Research Foundation. Part of this research was funded by the Australian Research Council (ARC)  Centre of Excellence for Gravitational Wave Discovery (OzGrav), CE170100004, and the ARC Centre of Excellence for All-sky Astrophysics (CAASTRO), CE110001020. J.C. acknowledges the ARC Future Fellowship grant FT130101219. Research support to I.An. is provided by the Australian Astronomical Observatory (AAO). J.M. is supported by the Hundred Talent Program, the Major Program of the Chinese Academy of Sciences (KJZD-EW-M06), the National Natural Science Foundation of China 11673062 and the Oversea Talent Program of Yunnan Province. M.G. thanks the Polish NCN grant OPUS 2015/17/B/ST9/03167. Some of the observations reported in this Letter were obtained with the Southern African Large Telescope (SALT) under proposal 2017-1-DDT-009. This research used resources of the National Energy Research Scientific Computing Center, a DOE Office of Science User Facility supported by the Office of Science of the U.S. Department of Energy under Contract No.~DE AC02-05CH11231. This research has made use of the NASA/IPAC Extragalactic Database (NED), which is operated by the Jet Propulsion Laboratory, California Institute of Technology, under contract with the National Aeronautics and Space Administration.

\bibliographystyle{aasjournal}

\end{document}